\documentstyle[a4,12pt,epsf]{article}

\begin{document}
\def\bra{\langle}
\def\ket{\rangle}
\def\ie{{\it i.e.}, }
\def\pref#1{(\ref{#1})}

\begin{center}
{\bf PHENOMENOLOGICAL THEORY OF FRICTION IN THE QUASISTATIC LIMIT :
COLLECTIVE PINNING AND MEMORY EFFECTS}

\begin{tabular}[t]{c}
Lyd\'eric Bocquet $^{1,2}$ and Henrik Jeldtoft Jensen $^{2}$\\
$^{1}$ Laboratoire de Physique, URA CNRS 1325 \\
E.N.S. Lyon, 46, all\'ee d'Italie \\ 69364 Lyon Cedex 07, France \\
$^{2}$Department of Mathematics, Imperial College\\
 180 Queen's Gate, London SW7 2BZ\\
United Kingdom
\end{tabular}
\end{center}

\begin{abstract}

When a elastic body is moved quasistatically back and forth over a surface, the
friction
of the interface is experimentally observed to circulate through a hysteretic
loop.
The asymptotic behaviour of the hysteresis loop is approached exponentially.
We describe how this behaviour is connected to the collective properties of the
elastic instabilities suffered by
the elastic body as it is displaced quasistatically.
We express the length scale of the exponential
in terms of the elasticity of the surface and the properties of
the rough substrate. The predicted scaling are confirmed numerically.
\end{abstract}

\def\abstractname{R\'esum\'e}
\begin{abstract}
Lorsqu'un mouvement quasistatique alternatif est impos\'e \`a un solide
\'elastique
en contact avec un support rugueux, il a \'et\'e mesur\'e exp\'erimen\-talement
que
la force de friction d\'ecrit une boucle d'hyst\'er\'esis. De plus, l'approche
de la
force de friction vers sa valeur asymptotique est exponentielle. Nous
d\'ecrivons
comment ce comportement est li\'e aux propri\'et\'es collectives des
instabilit\'es
\'elastiques subies par le solide lorsqu'il est d\'eplac\'e quasistatiquement.
La longueur
de relaxation de l'exponentielle est exprim\'ee en terme de l'\'elasticit\'e de
la surface
et des propri\'et\'es du support rugueux. Les lois d'\'echelle pr\'edites
sont confirm\'ees num\'eriquement.
\end{abstract}

\noindent {\bf short title} : Phenomenological study of quasistatic friction\\
\noindent submitted to Journal de Physique I

\noindent (1) author for correspondance : lbocquet@physique.ens-lyon.fr \\
\indent FAX : 0033 4 72 72 80 80
\newpage

\section{Introduction}

The present paper presents a discussion of the hysteretic behaviour of the
friction force
between two solid bodies. We use a one dimensional model to develop a
phenomenological analysis of friction in the quasi-static limit (zero velocity,
\ie {\it below} the
depinning threshold). In particular, we emphasize
the connection between {\it macroscopic} friction and
the {\it microscopic} elastic
instabilities occuring when an elastic body is pulled over a microscopically
rough surface.
Guided by
computer simulations of a one dimensional elastic chain, we construct a
coherent
collective pinning picture of friction. In particular, an equation for
the relationship between the friction force and the number of elastic
instabilities is
derived.
We use this equation to  obtain the hysteretic relation between displacement
and
friction force. Our model is qualitatively in agreement with experiments on
elastic friction \cite{Crassous}.

\subsection{Some open questions in friction}

It is useful to divide the problem concerning the friction between two
solid surfaces into a number of  sub-problems. At the microscopic quantum
level the forces have their origin in the electronic potential across the
interface between the two surfaces. These forces can be measured by various
types of scanning probe microscopy and can in some cases be calculated
by use of a quantum theory for the surface atoms of the solids \cite{Atomic}.
Microscopically
the two surfaces may often be considered as ideal periodic structures and
the interaction across the surfaces will be spatially homogeneous and periodic.

On the other hand, when two {\it macroscopic} surfaces are in contact, the
forces between the
surfaces will be due to the roughness of the surfaces on
macroscopic length scales. The friction force will originate in the
interlocking of a number of
asperities on the two surfaces.  If we can neglect plastic
deformations (tear and wear) of the asperities as they are forced past each
other, we   may expect that  it is possible to reduce part of the pinning
problem to
the interaction between  macroscopic elastically deformable asperities. We
can then neglect the particular quantum nature of the microscopic interaction
and attempt to establish an effective phenomenological theory of  the friction
between macroscopic elastic surfaces.  New controlled experiments on
artificially
constructed elastic surfaces make it urgent to understand theoretically this
model
system of friction \cite{Crassous}. Even this vastly reduced problem contains
some rather subtle
open questions.

Recently most of the theoretical and numerical works on the pinning problems
have focused on the
dynamical properties of friction in order to obtain  the friction-velocity
relationship {\it above the depinning transition}, \ie when the applied force
is
greater than the critical pinning force \cite{Dynamics1,DRG,Natterman}. In
particular dynamical
Renormalization Group techniques have been successfully applied in the low
velocity
limit by treating the depinning transition as a dynamical critical phenomena
\cite{DRG,Natterman}.

On the other hand, the litterature on the properties of pinning {\it below the
depinning transition} is mainly restricted to the pionneering work of Larkin
and
Ovchinnikov (LO) in the 70's \cite{Tinkham}. Their analysis first provided a
sketch of the
collective properties of pinning in order to construct the critical pinning
force.
Their predictions were successfully compared to experimental results on the
pinning
of vortex lines in superconductivity \cite{Tinkham}.
This success has led some authors to consider the behaviour below the critical
threshold
as essentially solved by LO \cite{Natterman}. However, in view of
new experimental results on quasistatic friction and contact line motion
\cite{Crassous,Crassous3},
it has become clear that
the LO analysis is incomplete on the following
important points. Firstly the LO argument does not involve {\it hysteretic
behaviour}.
Secondly their approach does not take any {\it mechanical instability} into
account,
while it has been shown that friction cannot arise without this crucial
ingredient
(see \cite{Nozieres,Jensen-1} and the discussion below). Moreover, new
experiments
have shown the existence of {\it memory effects} in the approach to the
critical pinning force,
which are characterized by a length scale \cite{Crassous,Crassous3,Heslot}.
In the experiments of Heslot {\it et al.} \cite{Heslot}, the
introduced length accounts for the ``age'' of the pulled system, while in the
experiments of Crassous {\it et al.} on friction and contact line motion
\cite{Crassous,Crassous3},
this length characterizes the
distance needed by the system to reach the stationnary state. Therefore it is
clear
that although the LO analysis captures an essential part of the collective
mechanism of pinning, the link to the above mentioned experimental observations
is still
lacking. The
construction of this link is the main object of the present work.

\subsection{The crucial role of elastic instabilities}

As already pointed out by Tomlinson \cite{Tomlinson}, friction arises in an
elastic media
due to the existence of a {\it multiplicity of metastable states}. This
multiplicity
induces mechanical instabilities in the system and
leads to dissipation and hysteresis \cite{Nozieres,Sokoloff,Jensen-1}.

The instabilities arise whenever the
local force-balance equation describing  the interlocked surfaces becomes
multivalued.
As the mutual displacement of the two bodies is increased, the asperities are
forced
against each other. The  interface force increases linearly as a function of
the centre
of mass displacement. When the local force balance becomes unable to sustain
the
mutual force between the asperities, a local rearrangement of the atoms in the
interface
will take place. If the roughness of the interacting surfaces is sufficiently
large, the
motion of the interface atoms takes place in the form of \underbar{swift} jumps
from one metastable
configuration to another. During such an instability the friction force drops
precipitously
by a certain value, for then  again to increase linearly upon further increase
in the relative
displacement of the two bodies.

Let us recall the microscopic features of the elastic instabilities.  This is
most easily
done by use of a single degree of freedom picture\cite{Jensen-1,Nozieres}.
Consider a particle
at position $x$  elastically coupled to a position $X$ (one may think of $X$ as
the
center of mass of  the elastic lattice). We want to follow the motion of the
particle as it
passes over an asperity modeled by a Gaussian peak in the potential energy. The
energy
of this system is given by
\begin{equation}
U= {1\over2}\kappa (x-X)^2 +A_p\exp[-(x/R_p)^2].
\end{equation}
The static equilibrium of the system is obtained by solving the equation
$\partial U/\partial x=0$ for a prescribed $X$ value, \ie we have to solve the
equation
\begin{equation}
\kappa (x-X) = {2A_p x\over R_p^2} \exp[-(x/R_p)^2].
\label{force-balance}
\end{equation}
This equation is solved graphically in Fig. \ref{force-fig}. The important
point is that
the equation has a single valued solution for any value of $X$ as long as the
slope
of  the force exerted by the Gaussian peak is everywhere smaller than $\kappa$.
This condition is
\begin{equation}
{2 A_p\over R_p^2} <\kappa.
\label{condition}
\end{equation}
When this condition is not fulfilled (as in Fig. \ref{force-fig}), the
equilibrium
position of the particle $x(X)$  will become a discontinuous function of $X$ :
a
mechanical instability occurs when the system passes over the defect.
Accordingly,
energy is dissipated (e.g. into the rapid degrees of freedom, like phonons) :
instabilities
induce hysteretic behaviour. However, this single particle mechanism does not
lead to memory effects. Therefore, in a more general case where an elastic
medium slides
over many defects, one still expects the basic mechanism for friction to be
linked
to the occurence of elastic instabilities; on the other hand, memory effects
are the indication
of the collective character of instabilities.

\subsection{Our aims }

Our aim in this paper is to give a mainly qualitative picture of collective
pinning
in the quasi-static limit in order to understand the memory effects. We stress
the
fact that as in the experiments of ref. \cite{Crassous}, we study the system in
a
regime {\it below the depinning transition} : the
sliding velocity of the two solid bodies strictly vanishes and no dynamical
effects are expected
in our case.
In particular the system will be considered at equilibrium under an applied
constraint.
The velocity dependence of the friction force is therefore not the purpose of
the
present study \cite{Dynamics1,DRG,Natterman}.

We will attack the problem along the following line :

(i) First we perform numerical simulations of a very simple model of the
system, in order
to check if the latter is able to reproduce the experimental results of ref.
\cite{Crassous}.
In particular, the following experimental facts should be recovered: existence
of an
{\bf hysteresis loop} when the system is pulled back and forth over the rough
surface; existence
of a {\bf memory length}; {\bf exponential approach} towards the stationnary
friction force.

(ii) Then we construct a ``microscopic'' scenario of friction by analysing
the microscopic behaviour of the system.

(iii) Finally, on the basis of the numerical results, we propose a qualitative
phenomenological model for the collective properties of friction in the
quasi-static
limit. This model provides the link between the collective {\it equilibrium}
and
{\it out-of-equilibrium} properties of the system below the depinning
threshold.

\section{General features of the numerical results}

\subsection{A simple numerical model}

Let us first introduce the model we use for our numerical experiments. For
simplicity we  model the two interacting elastic surfaces by a mobile
deformable
elastic medium in contact with a stationary undeformable rough surface.
This is clearly a limitation in comparison to two deformable elastic surfaces.
However, our findings will justify our anticipation that this simplification
is inessential.

Our model consists of a one dimensional string of particles at positions
$x_i$ where $i=1,...,N$, coupled together   through  elastic springs  all
of the same spring constant $k$. The chain, of length $L$, is assumed to be
periodic. We use the equilibrium length $a$
of the elastic springs as  our unit of length $a=1$.  The particles interact
with a set of pinning centres in the form of randomly positioned (at positions
$x^p_i$ where $i=1,...,N_p$) repulsive asperities of density $n_p$. All the
asperities are modelled by the same Gaussian potential peak of amplitude
$A_p$ and range $R_p$. The potential energy of the system can accordingly
be written as
\begin{eqnarray}
U &=& U_{el}+U_{pin}\nonumber\\
     &=&  {k\over 2} \sum_{i=1}^{L-1} (x_i-x_{i+1}-a)^2+\nonumber\\
     &&    \sum_{i=1}^L\sum_{j=1}^{N_p} A_p~\exp[-(x_i-x^p_j)^2/R_p^2].
\label{potential-energy}
\end{eqnarray}

We are interested in experiments where dynamical effects can be neglected.
For this reason we investigate the total force produced by the asperities
when the elastic chain is moved {\em quasistatically} through the asperities.

The simulations are performed in the following way. We start from the ideal
lattice configuration of the elastic chain. We then use molecular dynamics
annealing to relax the chain to the substrate potential. This is done by
starting the
particles out with a random distribution of  velocities. The Newtonian equation
of motion derived from the potential in Eq. \pref{potential-energy} is then
integrated
by the leap-frog algorithm \cite{Jensen-1}. While integrating the equation of
motion
 we gradually extract  kinetic energy by scaling the velocities of the
particles. In
this way we move the system towards one of the metastable configurations of low
potential energy.

When this initial relaxed configuration has been prepared we study the friction
(or ``pinning'') force induced when the centre of mass (c.o.m.) of the chain is
gradually
forced through the asperities at zero velocity $V\equiv 0$.
The most accurate way of numerically simulating the
quasistatic displacement of the c.o.m., $X$, is by performing cycles of shift
and relaxation.
The elastic chain is displaced as a {\em rigid} body a small  amount $dx$
($dx/a=10^{-5}$ in our
simulations)
by replacing all the particle positions by $x_i\mapsto x_i+dx$.  While the
c.o.m. is kept fixed at this new c.o.m. position $X_{new}=X_{old}+dx$,  the
chain is
next relaxed to the asperities. The c.o.m. is kept fixed by always
counteracting
the force exerted by the substrate by an external force $f_{ext}$ applied
homogeneously to  all the particles of the chain. We have
\begin{equation}
f_{ext}={1\over L} \sum_{i=1}^L {\partial U_{pin}\over\partial x_i}
\end{equation}
where $f_{ext}$ used in the present time step is calculated from the positions
of the
previous molecular dynamics  time step.  This method allows one to follow as
accurately
as desired the motion through the background potential. One could of course
also
move the chain by applying an external force larger than the present friction
force.
This method  leads to results qualitatively equivalent  to the above described
method.
The draw back of  simply applying a constant external force is that the force
from the substrate fluctuates in space. A constant force will therefore
sometimes
be much larger than the pinning force. This leads to unwanted acceleration
effects and
 make it difficult to remain in the limit of quasistatic motion
\cite{Jensen-4}.

\subsection{Existence of hysteresis}

As in the experiments of ref. \cite {Crassous}, we move the elastic chain back
and forth
over the ``rough'' surface, according to the numerical algorithm  discussed
above. For each
c.o.m. position of the chain, $X_{c.o.m.}$, we measure the force acting on the
particles of the elastic chain. This force will be denoted as the friction
force, $F_f$.
The whole curve $F_f$ vs. $X_{c.o.m.}$ is then averaged
over many cycles and over different random spatial configurations of pinning
centres.
Fig. \ref{averaged-force} shows a typical numerical result. Starting from a
given
point, the friction force reaches {\it after a finite distance} a plateau
value, independent
of the c.o.m. position. This plateau value is the {\bf static friction force} :
it is the
maximum value for an external force before the solid body (here, the elastic
chain)
aquires a non vanishing velocity.
Then, when the system is moved in the other direction, the same plateau value
with the
opposite sign is reached, but following a different curve in the $F_f$ versus
$X_{c.o.m.}$
plane : an hysteresis loop is
performed during a cycle.

As in the experiments, memory effects are clearly observed, since it takes a
finite
{\it length} for the system to reach the stationnary pinning force. This
defines
a memory length.

\subsection{Exponential decay of the pinning force}

As shown in Fig. \ref{exp-decay}, the approach towards the plateau value for
the friction force is \underbar{exponential}. For example, when the system is
pulled
in the $X_{c.o.m.}>0$ direction, the friction force can be very well fitted
after an initial small transient distance, by the following relation :
\begin{equation}
F_f(X_{c.o.m.})=F_{\infty}+ (F_0-F_{\infty})~\exp\left\{-{X_{c.o.m.}-X_0 \over
\zeta}\right\}
\label{Ff-exp}
\end{equation}
where $F_{\infty}$ is the plateau value for the force, $F_0$ is the force
measured at
a given point $X_0$. This relation defines the memory length, $\zeta$. In our
simulation, the latter was obtained to be of the order of a few $R_p$, the
range of
the defects.
The size of the initial transient regime was obtained to decrease when the
density of pinning centres $n_p$ increases.

This exponential approach towards the plateau value of the pinning force is in
perfect
agreement with the experimental results of ref. \cite{Crassous}. Therefore, the
simple
numerical model we use should contain the essential of the underlying physics
leading
to the memory effects and exponential decay. We can now analyse in greater
details
the simulation results to give a microscopic picture leading to these results.

\section{A first empirical understanding}

\subsection{Hooke's law and instabilities}

While in Fig. \ref{averaged-force}, we considered an average of the friction
force over many
different initial states, we now focus our attention on a particular
realization
of the numerical experiment. A typical non-averaged plot of the friction force
as a function of the c.o.m. displacement is shown in Fig.
\ref{non-averaged-force}. This figure is
characterized by a saw-tooth behaviour, which splits up into {\it linear
increase}
of the friction force, separated by {\it steep decrease} in the friction force.

(i) \underbar{The linear behaviour} corresponds to the {\bf reversible} linear
(elastic)
response of the system, when an external constraint
is applied in order to impose a given c.o.m. displacement.
This response is characterized by an elastic susceptibility,
$\kappa_L$, defined by
\begin{equation}
dF_f=\kappa_L \cdot dX_{c.o.m.}
\label{def_labusch}
\end{equation}
relating the measured infinitesimal change in the friction force, $dF_f$, to
the
c.o.m. displacement $dX_{c.o.m.}$, according to a simple Hooke's law. The
parameter $\kappa_L$
was first introduced by Labusch \cite{Labusch}, in the context of lattice
deformations
in crystals and will be denoted as the ``Labusch parameter'' in the following.

(ii)  \underbar{The discontinuities} in the force are the indication of a
dramatic
{\bf irreversible} transformation occuring in the system. A look at the
microscopic
trajectories shows that these jumps in the force are intimately connected to
a large {\it common} displacement of a significant number of particles
(typically of order 10-50
over 500).
The typical displacement $\delta x_0$ of each particle of the moving block  was
always of the
order of one
lattice spacing in all our simulations : $\delta x_0 \simeq a$.
This indicates that, because of the imposed c.o.m. displacement, the local
equilibrium state
of the system becomes unstable and a finite jump occurs towards a new local
equilibrium
state \cite{Jensen-1}. In other words, these jumps occur because the system
cannot support
elastically the applied external constraint anymore, so that a new metastable
state has
to be found in order to release the stored energy. Dissipation occurs during
these
abrupt transitions.
In the following, these jumps will be denoted as ``elastic instabilities''.

The previous scenario is in fact very reminiscent of the single particle case
(\ie
one spring over one defect), discussed above
\cite{Nozieres,Jensen-2}. In particular,
the multistability of the metastable postions seems thus to be recovered in the
many-particle
system. However
the characterization of the associated instabilities is now much more
difficult, because
of the collective character of the phenomena, as will be shown in section 4.

Let us assume that the elastic chain is pulled in the $X_{c.o.m.}>0$ direction.
We introduce
the spatial frequency of instabilities, $\nu$, defined as
the number of instabilities occuring in the system per unit c.o.m.
displacement; and
$\Delta F_{inst}$ the drop in force occuring in a single instability. It is
useful to
{\it define} it to be the absolute value of the drop, so that $\Delta
F_{inst}>0$.
The latter is assumed,
of course, to depend on the c.o.m. position (or equivalently on the friction
force)
at which the instability takes place. According to the
previous scenario, a simple differential equation for the (averaged) friction
force, $F_f$,
can be written :
\begin{equation}
{d F_f \over {dX_{c.o.m.}}} = \kappa_L - \nu ~\bra \Delta F_{inst} \ket
\label{equ_Ff}
\end{equation}
which accounts for both the elastic response of the system and the finite
change in the
force during instabilities. The notation $\bra \dots \ket$ means a
non-equilibrium
average over many different distributions of pinning centres {\it for a given
c.o.m. displacement},
or equivalently {\it for a given friction force}. We omit the brackets for
the Labusch parameter $\kappa_L$, since the latter is found in the simulations
to be independent of
the friction force $F_f$. This means that the elastic susceptibility of the
system
remains constant, when the friction force increases, \ie when new metastable
states
are explored. This rather astonishing fact will be discussed briefly in section
5.

We can see in Fig. \ref{non-averaged-force}
 that no instability occurs until
the force has reached the value $F_f =0$ and a rather long elastic relaxation
takes place up to that point : then $\nu=0$ for $F_f < 0$. In fact, this
is to be expected since in our numerical procedure
(involving an alternative forward/backward displacement of the whole system
like
in the real experiments), the initial state is {\it not} an equilibrium state
but the
stationnary state of the system when it is pulled in the opposite direction.
Therefore, until
some stress is {\it effectively} supported by the system (\ie $F_f>0$), the
system
relax elastically the external constraint and no instability occurs.
For $F_f > 0$, the instability process is turned
on. The  frequency of instabilities is found numerically
to reach a plateau value in a very short distance (much smaller than the
lattice spacing, $a$),
followed by a small decrease towards its stationnary value. This small
$F_f$ dependence will be omitted in the following and the frequency will be
considered
as roughly constant (for $F_f>0$).

\subsection{A phenomenological law and the exponential decay}

The averaged change in force during an instability,
$\bra \Delta F_{inst} \ket$ does not
remain constant when the c.o.m. is displaced. This can be seen for example in
Fig
\ref{non-averaged-force},
where the non-averaged friction force  is plotted versus the c.o.m.
displacement. The change
in force during an instability is seen to increase when the system is pulled
quasi-statically,
so that $\bra \Delta F_{inst} \ket$ is expected to increase when the friction
force $F_f$
increases. We conjecture a simple {\bf linear relationship} between these
two quantities :
\begin{equation}
\bra \Delta F_{inst} \ket = \delta F_0~+~\alpha~F_f
\label{DF_F}
\end{equation}
where $\delta F_0$ and $\alpha$ are two phenomenological parameters

This ``phenomenological law'' has been checked in the simulations by plotting
the averaged
change in an instability as a function of the friction force measured just
before the
instability takes place. Numerically, this procedure involves a simple
algorithm which
detects the instabilities. The latter relies on the measure of the numerical
derivative of the
friction force, which exhibits a dramatic change during an instability. This
rough indicator
has been checked to work with a very good accuracy. We have then averaged the
plot over
many different realizations of the initial conditions to compute {\it for a
given friction
force}, the corresponding averaged change in force $\bra \Delta F_{inst} \ket$.
A typical result
is plotted in Fig. \ref{phenom-law}. Except in the large force region, the
numerical points can
be fitted with a good agreement by a straight line, thus validating the
phenomenological relation
in eq. \pref{DF_F}. Physically, the increase of the force release during an
instability is
understandable. Roughly, when the external constraint increases, the
``susceptibility''
of the system increases accordingly, so that the drop in force during an
instability
becomes larger. This point will however be studied in more details in
the next section. The increase of $\bra \Delta F_{inst} \ket$ for large forces
in Fig.
\ref{phenom-law} could be the indication of the onset of large fluctuations
arising
in the close vincinity of the depinning threshold, although ``critical''
avalanches
(\ie involving the whole system) were not observed in our simulations.  But
this problem
requires a specific careful numerical work as done by Pla and Nori \cite{SOC}.
This is not the object of the present work.

Combining the phenomenological relation eq. \pref{DF_F} with eq. \pref{equ_Ff}
leads to
a {\it closed equation} for the friction force :
\begin{eqnarray}
{d F_f \over {dX_{c.o.m.}}} &= \kappa_L - \nu ~\bra \Delta F_{inst} \ket \\
&= \left(\kappa_L -\nu \delta F_0\right) - \nu \alpha ~ F_f
\nonumber
\label{eqn_Ff_2}
\end{eqnarray}
This equation predicts an {\bf exponential relaxation} of the friction force
towards a stationnary value given by
\begin{equation}
F_f^{\infty} = {\kappa_L \over {\nu\alpha}} - {\delta F_0 \over \alpha}
\label{Ff_stat}
\end{equation}
The relaxation length, $\zeta$, is related to the phenomenological parameter
$\alpha$, through the simple relation :
\begin{equation}
\zeta= 1/\nu\alpha
\label{mem-length}
\end{equation}
This exponential behaviour
is in agreement with both the numerical results (see Fig. \ref{averaged-force})
and the experimental
results of Crassous {\it et al.} \cite{Crassous}.

Therefore, the phenomenological relation eq. \pref{DF_F} provides an initial
``empirical''
understanding of the underlying physics of friction. However, how this simple
law
emerges from the microscopic picture remains to be clarified. Moreover, the
results obtained within this approach for the stationnary force (eq.
\pref{Ff_stat})
and relaxation length (eq. \pref{mem-length}) are only useful, if some
prediction
can be made
for their dependence on the physical parameters of the system ( e.g. density of
pinning centres $n_p$, strength and range of the pinning centres, $A_p, R_p$).
This
is the object of the next section.

\section{Microscopic picture : towards collective pinning}

\subsection{Description of instabilities}

In paragraph 3.1 we described the instabilities as an instantaneous collective
motion of an important number of particles of the system, while the other
remained
more or less stationary. Intuitively, it would be appealing to relate
the drop in the friction force to the number of jumping particles during the
instability.
To this end, we have studied numerically the microscopic behaviour of the
system
when an instability takes place. Especially, after isolating an instability, we
separated the system into the
block of "jumping particles" and the remaining particles. For both parts of the
system, we
computed the change in the force due to the defects. This operation was
repeated
for a number of instabilities of different amplitudes, taking place at
different c.o.m. positions.
The general ``rule'' we obtained was the following : (i) the
change in force measured for the jumping particles only was found to be {\it
very small};
(ii) the global drop in the total friction force observed in an instability is
caused
by the remaining particles, which didn't undergo a large displacement. This
can be interpreted as follows. When each of the $N_{jump}$ jumping particles
move {\it forward}
by a large, finite amount $\delta x_0$, the c.o.m. of the remaining particles
will
move {\it backward} by an amount $\Delta x_{c.o.m.} \simeq -~N_{jump}~\delta
x_0 / N$,
in order to keep the c.o.m. of the complete system constant. The corresponding
drop in force is then simply given by the relaxation of the corresponding
elastic constraint,
according to Hooke's law (see eq. \pref{def_labusch}) :
\begin{eqnarray}
F_{after}-F_{before} \equiv -\Delta F_{inst} &= \kappa_L~ \Delta x_{c.o.m.}
\nonumber \\
&= -~ {\kappa_L~\delta x_0 \over N}~N_{jump}
\label{DF_inst}
\end{eqnarray}
Eq. \pref{DF_inst} provides the desired link between the drop in the force
occuring in an instability
and the corresponding ``length'' of the instability, \ie the number of
particles that jumped, $N_{jump}$.

\subsection{Existence of a coherence length}

The afore-mentionned point (i) - that the change in force measured for the
jumping
particles only is extremely small - is very striking. In fact, this can be
considered as
a first indication of the {\em collective character of the instabilities}. To
understand this
subtle point, we first recall how an instability occur in the single particle
case,
as discussed by Nozi\`eres and Caroli \cite{Nozieres,Jensen-1}. The system
involves then only a
single particle attached to a spring and interacting with a defect. The
extremity of
the spring is moved adiabatically. For a sufficiently soft spring (see section
1.2
for details), the equilibrium position of the particle becomes multi-valued
over
a given range of system positions, \ie the system becomes unstable. The
evolution of the
particle thus exhibits three phases when
it encounters the defect :
first an adiabatic move, corresponding to the elastic response of the defect;
then a
large finite jump towards a point where the reaction force is very small : this
corresponds to the elastic ``instability''; finally, an
adiabatic move again, where only the tail of the defect's potential is felt by
the
particle. Accordingly, the drop in the force $\Delta f$ during the instability
is of the order
of the threshold force before the jump. Now consider $N_{jump}$ particles
jumping
according to such an ``individual scenario'', the global change in force for
the whole
system would simply be $\Delta F_{inst}\simeq N_{jump}~\Delta f$, which is
proportional
to the number of particles involved in the instability. This assumes that the
drop in the force
originates from the particles involved in the instability. But {\bf this
differs dramatically} from our observations in the numerical experiments.
Indeed according
to the previous points (i) and (ii), the crucial difference is that in our
simulated
systems, the drop in the force during an instability is only induced by the
release of the
elastic response of the particles {\em not involved} in the instability.

We interprete this point by assuming that the system separates into two types
of particles :
a few particles interact ``strongly'' with the defects, while all the other
only
interact weakly with the defects, \ie they can be considered as ``strongly
correlated''.
The equilibrium positions of one (or a few)
strongly pinned particle may become unstable, the latter thus performing a
large finite jump.
All
particles ``strongly connected'' to this particle then simply follow
collectively
the jump of the first jumping particle. Within this scenario, the change in
force
due to the ``jumping part'' of the system has no reason to be large. Indeed,
most of the
jumping particles (\ie except for the one -or the few- strongly ``pinned'')
do not interact strongly with the defects, so that their individual
change in force during their jump is mostly random, justifying the observation
facts
(i) and (ii).

There is another indication of the ``collective'' character of the
instabilities. Let us
come back to the phenomenological relation \pref{DF_F} for the drop in the
force during an instability, verified numerically on
Fig. \ref{phenom-law}. As can be observed in this figure, the value at the
origin,
$\delta F_0=\Delta F_{inst}(F_f=0)$, is non-vanishing. This means that, as soon
as
instabilities are allowed, a finite (non-zero) number of particles are involved
in the instability. Therefore, the length scale of the previously discussed
``strongly connected''
particles is non vanishing at equilibrium, and is hence an intrinsic property
of
the system at equilibrium.

These observations leed us to the notion of ``strongly
pinned'' particles and ``strongly connected'' particles, the latter being
characterized by a well defined correlation length. In fact, this separation
is reminiscent of the analysis introduced by Larkin and Ovchinnikov (LO)
\cite{Larkin,Tinkham},
of the pinning of vortex lines in superconductors. Their analysis is based
on the introduction of a ``correlation length'', characterizing the balance
between
the elastic energy (of the Abrikosov lattice in superconductors) and the
pinning
energy due to the interaction with the defects. This length can be equivalently
interpreted as
the displacement correlation length of the system.
More precisely, the linear size
of the correlated volume $l_L$ is defined as the length scale over which the
elastic rigidity of
the elastic medium is sufficient to counteract the random forces. The relative
shear
induced by the random forces is small within a correlated volume. The particles
within
a correlated volume are strongly correlated in the sense that if a particle is
displaced
a small amount, this displacement will be transmitted by the rigidity of the
lattice out to
distances of the order of the size of the correlated volume. At distances
larger than the
size of the correlated volume the random forces become essential. The
distortion of the
elastic lattice becomes appreciable and particles farther apart than $l_L$ are
only weakly
correlated. In this sense we can think of the elastic lattice as being broken
into weakly
interacting rigid subvolumes. It is at the interface between the correlated
volumes that
the elastic instabilities occur.

We now recall the (qualitative) LO argument in order to estimate $\ell_L$.

In the picture depicted above,
the total pinning force on a correlated volume, \ie the force due
to the defects, is the sum of many contributions of the same order $f_0\sim
A_p/R_p$, but
with {\it different signs} (since the strongly correlated particles only
interact
weakly with the pinning centres). Therefore, the
pinning force on a correlated volume involves a statistical summation over
all the individual contributions and will be of order $N_c^{1/2}~f_0$,
where $N_c=n_p~\ell_L$ is the number of pinning centres included in the
correlated volume
(we now restrict our study to a 1-D
problem).
Since there are $L/\ell_L$ correlated volumes in the system, the total pinning
force is found to be of order :
\begin{equation}
F_{pin} \sim L~\left( {n_p \over \ell_L} \right)^{1/2}~f_0
\label{Fpin}
\end{equation}
Since this force only acts for a distance of order $R_p$ (the range of the
pinning potential),
before changing randomly, the pinning energy is assumed to behave like
\begin{equation}
\delta {\cal E}_{pin}  \sim L~ \left( {n_p \over \ell_L} \right)^{1/2}~f_0\cdot
R_p
\label{Epin}
\end{equation}

On the other hand, the increase of elastic energy due to the deformation of the
lattice
inside a correlated volume can be estimated as $\delta {\cal E}_{el} = {1 \over
2} (k/n\ell_L)~\delta x^2$,
where $k$ is the bare spring constant, $n=1/a$ is the density of particles ($a$
is the
spring length), and $\delta x$ is the distortion distance. The term $k/n\ell_L$
takes
into account the {\it effective} stiffness of $n\ell_L$ springs contained in
a correlated volume. The distortion length is expected to be of order
of $R_p$, the range of the defect potential. For the whole system (involving
$L/\ell_L$
correlated volumes), we thus find
\begin{equation}
\delta {\cal E}_{el}  \sim L~{1\over 2}~k~a~\left(R_p\over \ell_L\right)^2
\label{Eel}
\end{equation}
Adding eqs. \pref{Epin} and \pref{Eel}, we find the change in energy per unit
length associated
with the randomly distributed defects :
\begin{equation}
\delta {\cal F} = {1\over 2}~k~a~\left(R_p\over \ell_L\right)^2 +
\left( {n_p \over \ell_L} \right)^{1/2}~f_0\cdot R_p
\label{Free_nrj}
\end{equation}

The optimized correlation length is found by assuming that the system evolves
towards
a state which minimizes the costs in energy. This is done
by minimizing the expression \pref{Free_nrj}
with respect to $\ell_L$, leading to
\begin{equation}
\ell_L = \left[ {2k~a~R_p \over {n_p^{1/2}~f_0}} \right]^{2/3}
\label{length}
\end{equation}
Replacing \pref{length} into eq. \pref{Fpin}, we obtain a total pinning force
which scales like
\begin{equation}
F_{pin} \sim  L\cdot{ n_p^{2/3} ~ f_0^{4/3} \over {k~a~R_p^{1/3}} }
\label{Fpin_scale}
\end{equation}

Larkin and Ovchinnikov \cite{Tinkham} assumed that the pinning force in eq.
\pref{Fpin_scale}
is {\bf the out-of-equilibrium pinning force}, defined as the force measured at
the point where the systems depins, \ie acquires a non-vanishing velocity. The
latter
corresponds to the symptotic force measured in our numerical experiments (or in
the
{\it real} experiments of ref. \cite{Crassous}) when the
system is pulled quasi-statically.

But our interpretation of $F_{pin}$ \underbar{differs} qualitatively from the
previous affirmation.
Indeed, the LO argument as depicted here is an {\bf equilibrium}
argument, which allows one to find the equilibrium properties of a lattice
interacting with randomly distributed defects. In this case, the averaged
pinning force should
vanish (otherwise the system is not at equilibrium). Therefore
the ``pinning force'' defined in eq. \pref{Fpin} cannot be identified with the
out-of-equilibrium pinning force. Moreover, the LO argument does not
take the elastic instabilities into account, while the numerical simulations
(and even
the real experiments, see \cite{Ciliberto} and \cite{DiMeglio}) show that these
instabilities play a crucial role in the pinning process.

In contrast to Larkin and Ovchinnikov,
we interpret $F_{pin}$ of eq. \pref{Fpin}, obtained along the previous
argument,
as a typical underlying scale for the {\it fluctuations} of the reaction force
due to the
defects at equilibrium : more precisely, it fixes the amplitude of the
root-mean-square
(r.m.s.) fluctuations of the pinning force. How this
scale is related to the non-vanishing, non-equilibrium pinning force (\ie the
plateau value
for the force for example in Fig. \ref{averaged-force}), remains to be
understood and will be the
object of the next sections.

\subsection{Linking the coherence length and the phenomenological relation}

In order to make the connection between the equilibrium properties of the
system
and our empirical understanding of friction depicted in section 3, we make the
following two {\bf conjectures} :

(A) in an instability, the number of particle undergoing a jump is given by the
length of the correlated volume. In other words, the instability is defined by
a block-jump of an entire correlated volume.

(B) when an external constraint is applied, the length of the correlated volume
\underbar{increases}.

Let us first discuss these two assumptions. The first one, (A), is intuitively
understandable.
Indeed, according to the picture already discussed in the previous paragraph,
instabilities are initiated by the few particles which are strongly interacting
with
the defects. When one of these strongly interacting particles becomes unstable,
and thus
peforms a large jump, all the particles which are strongly ``attached'' to this
jumping particle will follow as a whole.

The second assumption, (B), can be justified in the following way.
The correlation length, $\ell_L$ depends quite strongly on the
strength $A_p$ of the pinning centres, as can be seen in eq. \pref{length}
(remember
that $f_0\sim Ap/R_p$) : the weaker
the pinning centres are, the longer the correlation length is, because
the elastic lattice is then less distorted. Now, when an external force is
applied,
the {\it effective} strength of the defects will decrease, so that the
correlation length is
expected to increase when the system is moved, justifying conjecture (B).
To lowest order in the external force, we may expect a {\it linear} increase of
the correlation length, as would be obtained by applying linear response theory
to the system :
\begin{equation}
\ell_L (F_{ext})~=~\ell_L^{equ}~+~\gamma\cdot F_{ext}
\label{l_F}
\end{equation}
where $\ell_L^{equ}$ is the equilibrium correlation length obtained along the
LO argument, defined in \pref{length}. $F_{ext}$ is the
total external force applied to the system in the direction of the motion :
$F_{ext}={\bf F_{ext}}
\cdot {\mathaccent 94 {\bf e}}$, with ${\mathaccent 94 {\bf e}}$ the unit
vector pointing the direction of motion. In our case, the external force
applied in order to impose a given c.o.m. position simply identifies with the
measured
friction force, $F_f$, so that $F_f= F_{ext}$. Note that, at first sight,
considerations of
symetry would only predict a $F_{ext}^2$ dependence of $\ell_L$ as a function
of the
external field (because of the symetry $F_{ext} \rightarrow -F_{ext}$). However
such
a counter-intuitive linear behaviour usually occur in degenerate systems, where
a
careful perturbation theory has to be done \cite{Rae}. This is illustrated for
example
on the so called ``Stark effect'' for hydrogen, where a first order change in
the energy
of the first exited state (which is fourthfold degenerated) is found when an
electric
field is applied. In our case, the degeneracy stems from the multistability of
the
metastable states, which defines spatial ``multipoles''. The linear change in
$\ell_L$
may be thought as arising from the interaction between these multipoles and the
applied field. Moreover, the linear guess will be confirmed {\it a posteriori}
by the
consistency of the scenario which results from it.

By use of the two conjectures, we can now understand the ``phenomenological
relation'', eq. \pref{DF_F}, relating linearly the drop in force measured an
instability to the friction force. Combining eqs. \pref{l_F} and
\pref{DF_inst},
and setting $N_{jump}=n~\ell_L (F_f)$ according to the previous argument, we
obtain
\begin{eqnarray}
&\Delta F_{inst} &= {\kappa_L~n.\ell_L(F_f)~\delta x_0 \over N}
	\nonumber \\
& &= {\kappa_L~n.\ell_L^{equ}~\delta x_0 \over N}~+~
           \left({\kappa_L~n~\gamma~\delta x_0 \over N}\right)\cdot F_f
	\nonumber \\
\label{calcul}
\end{eqnarray}
The proposed scenario thus gives a coherent picture of the mechanism leading to
the
phenomenological law. Moreover,
identifying this expression with the phenomenological relation eq. \pref{DF_F},
we are left
with the following ``microscopic'' expressions for the two phenomenological
parameters $\delta F_0$ and $\alpha$ :
\begin{eqnarray}
&\delta F_0 = {\kappa_L~n.\ell_L^{equ}~\delta x_0 \over N} \nonumber \\
& \alpha = {\kappa_L~n~\gamma~\delta x_0 \over N}
\label{def}
\end{eqnarray}

One of the important
consequences of eqs. \pref{def} is that it will allow us to compute the
dependence
of the two introduced phenomenological quantities on the microscopic parameters
of the system (density of pinning centres $n_p$, strength and range of the
defects $A_p$, $R_p$).
This requires more information on the Labusch parameter $\kappa_L$, which will
be
obtained in the next section.

For the dimensionless parameter $\alpha$, the derivation is more subtle since
it
involves the new parameter $\gamma$ introduced in eq. \pref{l_F}. If we think
of
this parameter in terms of a susceptibility (as usually done in linear response
theory), then it is determined by the properties of the system {\it at
equilibrium} \cite{Forster}. Now, according to its definition \pref{l_F}, it
has
the dimensionality of a length divided by a force. But for a system of springs
interacting with randomly distributed defects, we expect the length scale to be
fixed
by the correlation length scale introduced in the LO argument
(see eq. \pref{length}), while
the force should be fixed by a typical value of the r.m.s. fluctuations of the
pinning
force. According to our interpretation of the LO argument, this scale is
fixed by $F_{pin}$, introduced in eq. \pref{Fpin}. Therefore we are left
with the following expression
for the  parameter $\gamma$ :
\begin{equation}
\gamma \sim {\ell_L^{equ} \over {F_{pin}} } \sim \ell_L^{equ}~ \left(
{\ell_L^{equ} \over
n_p} \right)^{1/2} ~ {1\over{L~f_0}}
\label{gamma}
\end{equation}
By replacing eq. \pref{gamma} into eq. \pref{def}, one obtains for
the phenomenological parameter $\alpha$ the expression
\begin{equation}
 \alpha \sim {\kappa_L\over N}~n~\delta x_0 \cdot {\ell_L^{3/2} \over{
L~f_0~n_p^{1/2} } }
\label{alpha}
\end{equation}
where the expression for $\ell_L$ is given in eq. \pref{length}. This
expression can be
written in a more transparent form :
\begin{equation}
\alpha \sim {\delta F_0 \over F_{pin}}
\label{alpha-2}
\end{equation}

The  predictions of eqs. \pref{def} and \pref{alpha} will be compared with the
numerical
results in section 5.3. This will allow us to assess the validity of our
approach.
But we first need to characterize more properly the dependence of the other
quantities,
in particular of the Labusch parameter. This is done now.

\section{Friction and scaling laws}

\subsection{The Labusch parameter}

The Labusch parameter is defined in eq. \pref{def_labusch} as the elastic
susceptibility
of the system. By definition, this susceptibility measures the response
of the pinning centres (defects) when the particles interacting through the
springs
are slightly displaced from their equilibrium positions, \ie when
the applied force balances the pinning force \cite{Jensen-4}.
We performed numerically this ``experiment'' by displacing
the c.o.m. by a very small amount and analysed the corresponding change in
force on each defect.
Surprisingly, the change in the reaction force of the defects was found to be
homogeneously distributed
over the defects, thus leading to a picture of an individual elastic response
of the
pinning centres to the diplacement of the spring system.

Therefore, we expect the Labusch parameter to scale like
\begin{equation}
\kappa_L= n_p L~k_0
\label{kappa}
\end{equation}
where $n_p$ is the density of pinning centres and $L$ the length of the system.
$k_0$
is the individual elastic response of a defect, and can be roughly estimated as
the
probability for a defect to interact with a particle, of order $\sim n~R_p$
($n$ being the
particle density and $R_p$ the range of the potential), multiplied by a typical
value of the
second derivative of the pinning potential, of order $A_p/R_p^2$. This leads to
\begin{equation}
k_0 \sim n~R_p~{A_p \over R_p^2}
\label{k0}
\end{equation}
Combining, eqs. \pref{kappa} and \pref{k0}, we obtain that the Labusch
parameter should
scale like
\begin{equation}
\kappa_L \sim n_p~A_p~R_p^{-1}
\label{scaling-Labusch}
\end{equation}
This scaling relation has been checked in the
simulations, by varying $n_p$ for a given set of potential parameters $A_p$,
$R_p$, and
varying $A_p$, $R_p$ for a given density. The numerical results are obtained to
be in
reasonable agreement with these predictions. The dependence on $n_p$ is shown
on
Fig. \ref{fig-scaling-npb} : the numerical value of $\kappa_L$ is plotted as
triangles;
the underlying dotted
curve is a straight line with slope $1$. We explored numerically
the dependence on the strength and range of the defect potential too.
This was done by varying slowly $A_p$ and $R_p$
for a given density. The presented results do not involve however an average
over
different random spatial distributions of the pinning centres. The measured
points for
$\kappa_L$ are plotted as circles in Fig. \ref{fig-scaling-Ap} for the
dependence on $A_p$ and in Fig. \ref{fig-scaling-Rp} for the dependence on
$R_p$ :
The dotted lines are a guide for the eye to indicate the predicted slopes :
slope $1$
for the dependence on $A_p$ (in Fig. \ref{fig-scaling-Ap})
and slope $-1$ for the dependence on $R_p$ (in Fig. \ref{fig-scaling-Rp}).
The trend is seen to be correct for both dependences.
However a more extensive numerical study is still
needed for a whole set of parameters $A_p$ and $R_p$, using different densities
of pinning centres.

It is interesting to note that the picture of an individual response of the
pinning centres
for the elastic susceptibility of the system is coherent with the fact that the
Labusch parameter is found in the simulations to be independent of the friction
force
(\ie of the applied constraint). We confirmed this result in all our
simulations, by checking
that the slope of the linear part of the non-average friction force, as in Fig.
\ref{non-averaged-force}, do not change when the c.o.m. of the system is
displaced.
Apart from the previous qualitative argument, based on an individual response
of the
pinning centres, this result remains up to now quite obscure to us.

\subsection{The Frequency of instabilities}

Finally, to complete our phenomenological description, we need to characterize
more precisely
the {\it frequency} of instabilities, $\nu$. Up to now we do not have a full
understanding
of the mechanism of {\it creation} of instabilities, but some predictions can
however
be made.

In the {\it stationary} regime, the following relation is obtained from eqs.
\pref{equ_Ff}
and \pref{calcul}
\begin{equation}
\kappa_L=\nu~\Delta F_{inst}=\nu~{\kappa_L~n\ell_L(F_f^{\infty})~\delta x_0
\over N}
\label{nu_stat}
\end{equation}
where $F_f^{\infty}$ is the plateau value of the friction force. This equality
imposes the
scaling of the asymptotic frequency $\nu$ to be
\begin{equation}
\nu \sim { N \over {n~\delta x_0}}~{ 1\over \ell_L^{equ}}
\label{scaling-nu}
\end{equation}
Note that here we are only
interested in the {\it scaling} properties of the frequency $\nu$, so that we
forget any
friction force dependence of $\ell_L$ to focus on its dependence on density of
defects and other
microscopic parameters :  we simply use the fact that the stationary value of
the correlation
length $\ell_L(F_f^{\infty})$ has the same scaling on $n_p$, etc...
as $\ell_L^{equ}$ defined in eq. \pref{length}.
We cannot a priori apply the expression for $\nu$ in eq. \pref{scaling-nu} to
the
transient regime.

However an alternative argument can be given, which estimates $\nu$ from the
fact that the system is broken
up into the weakly and strongly pinned particles.
Recall the mechanism producing the instabilities in the
single particle problem (see 1.2.) : a
particle P (position $x$) is elastically coupled to a position $X$ and interact
with a
defect. In the elastic region,
we may approximate the force due to the defect by a Hooke's law, with a
stiffness $k_d
\sim A_p / R_p^2$. Then, if the position $X$ is displaced by a small amount
$\delta X$,
the equilibrium position of the particle P will be displaced by an distance
$\delta x
\simeq \kappa~\delta X /(\kappa+k_d)$, where $\kappa$ is the spring constant.
If the stiffness of the
spring is much weaker than $k_d$ (which is a condition for the existence of
instabilities,
see e.g. \pref{condition}), we obtain $\delta x \sim \kappa ~ \delta X / k_d$.
Now in our many
body system, we can identify the position $X$ with the c.o.m. position
$X_{c.o.m.}$,
and the stiffness of the spring $\kappa$ with the effective stiffness of the
correlated
volume, $\kappa \sim k/(n\ell_L)$. Moreover the particle P can be identified
with one
strongly pinned particle, since in our microscopic scenario the latter are
expected to
induce the instabilities.
An instability will occur when the latter will move over a distance of order of
the
range of the potential : $\delta x \sim R_p$. This is equivalent to a change in
c.o.m. position
$\delta X$ given
by $\delta X \sim n \ell_L~k_d~R_p/k$, according to our single particle
discussion.
This argument thus predicts a frequency scaling like
\begin{equation}
\nu \sim {k \over{k_d~n~R_p}} {1 \over \ell_L^{equ}}
\label{nu}
\end{equation}
which is consistent -though not strictly identical- with our first guess, eq.
\pref{scaling-nu}.

In particular, this argument predicts a
dependence of the frequency of instabilities on the density of pinning centres
as
$\nu \sim n_p^{1/3}$ (see eq. \pref{length}). The numerical results are
consistent with this
scaling as shown in Fig. \ref{fig-scaling-npa} : the circles are the numerical
points; the
dotted line has a slope $1/3$.

\subsection{Summary of the predictions}

At this stage, we can check the predictions for the phenomenological parameters
and
measured quantities obtained within our scenario : in particular for the
two phenomenological paramaters $\delta F_0$ and $\alpha$, for the Labsuch
parameter,
the plateau value of the friction force $F_f^{\infty}$ and the memory length
$\zeta$. Apart from the pinning force and the Labusch parameter, we have mainly
focussed
our numerical study on the dependence of the measured quantities on the density
of pinning
centres $n_p$. Since the predicted scalings on $n_p$ are not obvious (see the
different powers of $n_p$ in the equations below !), the comparison with the
numerical
results as a function of $n_p$ are therefore expected to be already a drastic
test for our scenario.
But clearly more extensive numerical work needs to be done to explore the whole
parameter space ($n_p$, $A_p$, $R_p$).

First we focus on the phenomenological parameters, $\delta F_0$ and $\alpha$.
Their ``microscopic''
expressions are given in eqs. \pref{def} and \pref{alpha}.
Combining the predicted scaling of the Labusch parameter, as displayed
in eq. \pref{kappa} with the expression of the correlation length
$\ell_L^{equ}$, given in
eq. \pref{length}, we obtain the following dependence as a function of the
density of
pinning centres, $n_p$ :
\begin{eqnarray}
&\delta F_0 &\sim n_p^{2/3} \nonumber\\
&\alpha & \sim n_p^{0}
\label{scalings-phenom}
\end{eqnarray}
In particular, this shows that the slope $\alpha$ of the phenomenological law
eq. \pref{DF_F}
does not depend on the density of pinning centres !
Both predictions of eq. \pref{scalings-phenom} have been checked numerically.
The results
are plotted in Fig. \ref{dF0-alpha-scaling} : symbols represent the numerical
points and
the dashed lines, the predicted scalings. The agreement is seen to be correct
for both
parameters. This confirmation is crucial, since it shows that the proposed
scenario
for the underlying mechanism of friction (see section 4) is coherent. In
particular,
this justifies our two conjectures (A) and (B) of section 4.3.

Now we turn our attention to the other measured quantities.
According to eq. \pref{Ff_stat}, the stationary friction force $F_f^{\infty}$
is the
sum of two term, ${\kappa_L /{\nu\alpha}}$ and ${\delta F_0/\alpha}$. However
combining the ``microscopic'' expressions obtained for all the different
parameters involved,
eqs. \pref{def},\pref{alpha},\pref{alpha-2},\pref{scaling-nu}, we obtain that
{\bf the stationnary
friction
force $F_f^{\infty}$ simply scales like $F_{pin}$, the r.m.s. value of the
fluctutations in
equilibrium}. Using eq. \pref{Fpin_scale}, this leads to
\begin{eqnarray}
&F_f^{\infty}&\sim F_{pin} \sim L~{ n_p^{2/3} ~ f_0^{4/3} \over {k~a~R_p^{1/3}}
}
		\nonumber \\
&&\sim n_p^{2/3}~A_p^{4/3}~R_p^{-5/3}
\label{scaling-Ff}
\end{eqnarray}

On the other hand, using \pref{alpha},\pref{alpha-2},\pref{scaling-nu},
the memory length $\zeta$ is found to scale
like
\begin{eqnarray}
& \zeta & = {1 \over {\nu\alpha}} \nonumber \\
&       & \sim {F_{pin} \over \kappa_L} \sim n_p^{-1/3}~A_p^{1/3}~R_p^{-2/3}
\label{scaling-zeta}
\end{eqnarray}
where the microscopic expressions for $F_{pin}$ and $\kappa_L$ have been used
(see
eqs. \pref{Fpin_scale},\pref{kappa}).

In fact, an equivalent relaxation length can be introduced, defined as the
ratio
between the asymptotic friction force $F_f^{\infty}$ and the Labusch constant
$\kappa_L$ :
\begin{equation}
\delta \equiv {F_f^{\infty} \over \kappa_L}
\label{def-delta}
\end{equation}
Since $F_f^{\infty}$ scales like $F_{pin}$, this length exhibits the {\it same
scaling} as the
memory length $\zeta$ :
\begin{equation}
\delta \sim \zeta \sim n_p^{-1/3}~A_p^{1/3}~R_p^{-2/3}
\label{scaling-delta}
\end{equation}

On Figs. \ref{fig-scaling-npa} and \ref{fig-scaling-npb}, we summarize the
numerical results for
the scalings of the different
quantities as a function of the density of pinning centres. The symbols are the
numerical
points and the dashed lines indicate the predicted scalings. The errorbars were
estimated
in the following way : for a given (random) distribution of pinning centres,
the measured quantities
were averaged over many cycles (forward and backward motion), giving a
particular value
for the desired parameter. Then we averaged again on different random
configurations
of pinning centres : this gives a mean value and a typical errorbar for each
quantity.

For the pinning force and the Labusch parameter, the errorbars are within the
size of
the symbols and thus not displayed explicitly in the figure. For both
parameters, the
numerical results (triangles for $\kappa_L$ and crosses for $F_f^{\infty}$)
are in good agreement with the predicted scalings of eqs.
\pref{scaling-Ff} and \pref{scaling-Labusch} : these are illustrated by the
dotted lines in the
figure, with slope $1$ for $\kappa_L$ and slope $4/3$ for $F_f^{\infty}$.

Note however the large errorbars involved in the estimation
of the memory length, $\zeta$. Indeed, in contrast to the other parameters, the
estimated
value of $\zeta$ fluctuates a lot when different configurations
of pinning centres are considered. Good statistics would therefore involve an
average over
many different (randomly distributed) configurations of defects. Such a
procedure requires
much more numerical work and goes beyond the purpose of this first stage study,
devoted to
construct a coherent picture of friction. The numerical results for $\zeta$
(circles)
are however coherent with
the proposed scaling \pref{scaling-zeta}, as shown in Fig.
\ref{fig-scaling-npb}. The underlying
dotted line has a slope $-1/3$, as predicted by eq. \pref{scaling-zeta}.

On the other hand, the estimate of the other length $\delta$ is obtained to be
more precise,
since the errorbars involved in the numerical calculations of $F_f^{\infty}$
and $\kappa_L$ are
both quite small. The numerical results (squares) are seen to follow quite
closely the
predicted scaling of eq. \pref{scaling-delta}, illustrated in the figure by a
line
with a slope $-1/3$.

Finally we checked the dependence of the Labusch paramater and of the pinning
force
as a function of the strength $A_p$ and range $R_p$ of the defects, as given by
eqs. \pref{kappa} and \pref{scaling-Ff}. The results are summarized in Figs.
\ref{fig-scaling-Ap} and \ref{fig-scaling-Rp}.
The dependence of $\kappa_L$ has already been discussed in section 5.1. The
results
for the friction force $F_f^{\infty}$ are plotted as triangles in both figures.
The predictions of eq. \pref{scaling-Ff} are illustrated as dotted lines of
slopes $4/3$ for the dependence on $A_p$ and $-5/3$ for the dependence on
$R_p$.
Here again, the numerical results are obtained to be in good agreement with the
theoretical predictions. Note that the agreement of the measured $F_f^{\infty}$
and $\kappa_L$
with the predicted scalings on $A_p$ and $R_p$ implies necessarily that
$\delta$
follows the predicted dependence on $A_p$ and $R_p$ as displayed on eq.
\ref{scaling-delta}.

Let us emphasize again that we only performed a partial
exploration of the parameter space $\{n_p,A_p,R_p\}$, and a more complete
numerical
work is still to be done. However within the explored window,
the trend of {\it all} the quantities involved in the proposed scenario is
obtained to
follow the theoretical predictions. Therefore, the numerical results validate
with some
confidence the proposed picture for the onset of friction.

\section{Discussion and conclusions}

This work focuses on the onset of friction at the interface between an elastic
media
and a ``rough'' surface. In particular, we study how the
friction force evolves when the system is pulled {\it quasistatically} over the
surface.
Recent experiments have shown that, in this controlled limit, the system
exhibits
hysteresis and more surprinsingly, memory effects \cite{Crassous}. These
memory effects are characterized by a {\it length scale}.
While it was known for a while
that elasticy is able to produce hysteresis and dissipation, it is shown here
that the
memory effects originate from the {\it collective character} of the induced
elastic instabilities.

The collective character of quasistatic friction is related to the existence of
a
correlation length in the system {\it at equilibrium}, which expresses the
balance between
the elasticity of the
medium and pinning to the defects. When an external constraint is applied to
the system, this
length is shown to increase linearly with the applied force. This linear
increase provides
a simple explanation for the {\it exponential} approach to the stationary
static pinning force, as observed numerically and experimentally
\cite{Crassous}.
Moreover the evaluation of this length
allowed us to make some predictions for the dependence of the stationnary
pinning
force and memory length on the microscopic parameters of the system (density,
strength
and range of the defects of the surface). In particular, it is shown that the
value of
the {\it out-of equilibrium} stationary pinning force is fixed by the typical
scale for the {\it fluctuations} of the pinning force {\it in equilibrium}. The
link between
the two limits
is provided by the analysis of the increase of the correlation length out of
equilibrium
in the spirit of linear response theory. The predicted scalings are in
agreement
with the measured numerical results, validating therefore the proposed
scenario.

However, our study only focuses on the {\it quasistatic} motion of the elastic
media,
\ie when any dynamical effect can be neglected. When the system aquires a {\it
finite}
velocity, one may expect that the main lines of the proposed scenario will
subsist and could provide an interesting alternative route to {\it dynamical}
friction. Such
a hope is supported by the analogies between some of our results and the
heuristic
model of Heslot {\it et al.} \cite{Heslot}.  They show in particular that
experimental
results for sliding dynamics in the creep controlled regime can be accounted
for
by a simple model, based mainly on the assumption that the properties of the
system
depend on its ``age'', $\phi$, for which a ``successful'' definition was found
to be :
\begin{equation}
\phi \equiv \int_{t_0}^t \exp\left[ - {x(t)-x(t^{\prime}) \over D_0} \right]
{}~dt^{\prime}
\label{def_age}
\end{equation}
where $x$ is the c.o.m. position of the moving body and $D_0$ is a ``memory
length''.
In their work, $D_0$ is assumed to account for plastic flow,
but memory effects have been shown in this work to occur even in the elastic
limit.
This indicates that memory effects do subsist above the depinning threshold.
However
they have not been discussed on microscopic grounds up to now.
The striking analogy
between this definition of the age and
the exponential decay towards the stationnary friction force in our case may be
only fortuitous,
but this should anyway deserve some more careful investigation.
However, such a generalization
should be handled with care. In particular, as already emphasized in paragraph
5.2,
our understanding of the {\it creation} of instabilities is only qualitative.
In the dynamical
case (finite sliding velocity), sound waves induced by instablilities may play
a crucial
role. These effects are not included in our discussion of the quasistatic
limit.

Finally, it is interesting to note that our approach may be generalized to the
problem of contact line motion in the quasi-static limit. This problem
has been investigated experimentely very recently on nanoscales, using the
Surface Force Apparatus (SFA) technique \cite{Crassous3}. The experimental
results
present a striking analogy with those obtained in the experiments on dry
friction
in the quasistatic limit. In particular, the force acting on the contact line
(related
to the contact angle of the line) is shown to present hysteresis as a function
of the {\it displacement} of the contact line, defining spatial memory effects
characterized
by a length scale. Moreover, the approach to the stationnary value of the force
(\ie of the
advancing or receding contact angle) was obtained experimentaly to be {\it
exponential} too.
Obviously, all these
facts are very similar of those obtained both experimentally and numerically
for
dry friction. In the case of contact line motion, we expect the motion to be
controlled
by the competition between the pinning of the contact line to the defects of
the surface and on the other hand, the
``stiffness'' of the line in the direction perpendicular to the motion.
If we neglect the energy contribution  of the already ``wetted'' defects, the
quasistatic
motion of the contact line
reduces to a one dimensional problem and it is thus
tempting to apply our results.
This would lead to
a force acting on the contact line, $H$ (denoted as the ``hysteresis of the
contact line''
in the contact line language) which scales with  the number of defects $N_p$
as $H\sim N_p^{2/3}$ (see section 5). This guess is in rather good agreement
with both
numerical results of ref. \cite{Crassous2}, where a scaling $H\sim N_p^{0.7}$
was obtained,
and even with the experimental results of Di Meglio \cite{DiMeglio},
where a scaling $H \sim N_p^{0.8}$ was measured. Moreover in the latter
reference, it was
explicitly stressed that avalanches were playing a crucial role in constructing
the
hysteresis properties. These avalanches are very reminiscent of the
instabilities
observed in our numerical simulations. Beyond this encouraging agreement,
further work is needed
to understand more deeply these analogies and characterize more precisely the
regime
in which these analogies are relevant.

More generally, it is interesting to note that the {\it nature} of
instabilities
(here elastic instabilities) does not play a crucial role in the {\it
qualitative}
understanding of the onset of friction. The hope would then be to be able to
include {\it plastic
effects} into the scenario too, along the same lines as {\it elastic effects}.
Work along these lines is in progress.

{\it Acknowledgement}
The authors would like to thank J\'er\^ome Crassous and Elisabeth Charlaix for
communicating their experimental results prior to publication. L.B. is grateful
to Joergen Vitting Andersen for helpful discussions.
L.B. acknowledges support of a NATO grant for scientific research during his
stay at
Imperial College, London.

\newpage

{{\bf Figure 1} : Graphic resolution of
eq. \pref{force-balance} (in the multistable
case $2A_p/R_p^2 > \kappa$). The full line
is the derivative of the defect gaussian potential. The slope of
the straight line is $\kappa$, the stiffness of the spring. The intersection
gives
the solution of eq. \pref{force-balance}. Black dots are the stable solutions,
the open
dot is the unstable solution.
\label{force-fig}}

{{\bf Figure 2} : Hysteresis of the averaged force as a function of the
center of mass displacement $X_{com}$ when the elastic chain is pulled back and
forth
over the defect surface. The parameters are : length of the elastic chain
$N=500$;
density of pinning centres $n_p=0.5$;
parameters of the gaussian defect potential  $A_p=0.06$, $R_p=0.25$.
\label{averaged-force} }

{{\bf Figure 3} : Log-linear plot of the averaged force as a function of
the center of mass displacement $X_{com}$ for two different densities (top
curve $n_p=0.35$,
bottom curve $n_p=0.2$). The dotted lines are an exponential fit of the
averaged force in the large $X_{com}$ limit. The other parameters are $N=500$,
$A_p=0.06$
and $R_p=0.25$.
\label{exp-decay} }

{{\bf Figure 4} : Plot of the {\it non-averaged} force as a function
of the center of mass displacement, \ie for one realization of the cycle.
The parameters are $N=500$, $n_p=0.2$, $A_p=0.06$
and $R_p=0.25$.
\label{non-averaged-force} }

{{\bf Figure 5} : Average of the discontinuity in force during an instability,
$\bra \Delta F_{inst}\ket$ as a function of the friction force $F_f$. The
dotted line
is linear fit of the numerical points (the few last ones excepted). The slope
gives the
phenomenological parameter
$\alpha$, while its value at the origin gives $\delta F_0$. For the numerical
parameters
under consideration ($N=500$, $n_p=0.8$, $A_p=0.06$, $R_p=0.25$), this gives
$\alpha=0.095$,
$\delta F_0=1.3$.
\label{phenom-law} }

{{\bf Figure 6} : Log-Log plot of the two phenomenological parameters
$\alpha$ (squares) and $\delta F_0$ (circles), as a function of the
density of pinning centres $n_p$. The straight lines show the predicted
scalings : $n_p^0$
for $\alpha$ and $n_p^{2/3}$ for $\delta F_0$.
The length of the chain is $N=500$
and the parameters for the potential of pinning centres are $A_p=0.06$,
$R_p=0.25$.
\label{dF0-alpha-scaling} }

{{\bf Figure 7} : Log-Log plot of the frequency of instabilities as a function
of the density of pinning centres $n_p$. The straight line has a slope $1/3$.
The length of the chain is $N=500$ and
the parameters for the potential of pinning centres are $A_p=0.06$, $R_p=0.25$.
\label{fig-scaling-npa} }

{{\bf Figure 8} : Log-Log plot of the different measured quantities as a
function
of the density of pinning centres $n_p$. The triangles stand for the Labsuch
parameter $\kappa_L$; the crosses for the stationnary static pinning force
$F_f^{\infty}$;
the circles for the memory length $\zeta$; and the squares for the
length $\delta$. The straight lines indicate the scalings predicted within the
proposed scenario.
{}From bottom to top, the slopes of the lines are $-1/3$, $-1/3$, $2/3$, $1$.
The length of the chain is $N=500$ and
the parameters of the potential of pinning centres are $A_p=0.06$, $R_p=0.25$.
\label{fig-scaling-npb} }

{{\bf Figure 9} : Log-Log plot of the stationnary pinning force (triangles) and
of
the Labusch parameter (circles) as a function of the strength of the pinning
centres, $A_p$.
The straight lines indicate the predicted scalings : $F_f^{\infty} \sim
A_p^{4/3}$; $\kappa_L\sim A_p^{1}$.
The length of the chain is $N=500$ and the density of pinning centres
$n_p=0.5$.
\label{fig-scaling-Ap} }

{{\bf Figure 10} : Same as Fig. \ref{fig-scaling-Ap} but for the dependence on
the range of the
potential, $R_p$.
The straight lines indicate the predicted scalings : $F_f^{\infty} \sim
R_p^{-5/3}$; $\kappa_L\sim R_p^{-1}$.
The length of the chain is $N=500$ and the density of pinning centres
$n_p=0.5$.
\label{fig-scaling-Rp} }
\newpage

\end{document}